\documentclass[epj]{svjour}
\usepackage{graphicx}
\usepackage{bm}
\begin{document}

\title{Determination of the interactions in confined macroscopic Wigner
islands: theory and experiments}
\author{P. Galatola\inst{1} \and G. Coupier\inst{2} \and M. Saint
Jean\inst{2} \and J.-B. Fournier\inst{3,1} \and C. Guthmann \inst{2}}

\institute{Laboratoire Mati\`ere et Syst\`emes Complexes, UMR 7057 CNRS
\& Universit\'e Paris~7 - 2 place Jussieu, F-75251 Paris Cedex 05,
France \and
Laboratoire Mati\`ere et Syst\`emes Complexes, UMR 7057 CNRS
\& Universit\'e Paris~7 - 140 rue de Lourmel, F-75015 Paris, France \and
Laboratoire de Physico-Chimie Th\'eorique, UMR 7083 CNRS - ESPCI,
10 rue Vauquelin, F-75231 Paris Cedex 05, France}

\date{\today}

\abstract{Macroscopic Wigner islands present an interesting
complementary approach to explore the properties of two-dimensional
confined particles systems. In this work, we characterize theoretically
and experimentally the interaction between their basic components, viz.,
conducting spheres lying on the bottom electrode of a plane condenser.
We show that the interaction energy can be approximately described by a
decaying exponential as well as by a modified Bessel function of the
second kind. In particular, this implies that the interactions in this
system, whose characteristics are easily controllable, are the same as
those between vortices in type-II superconductors.}

\maketitle

\section{Introduction}
\label{intro}

The recent development of in-situ imaging techniques has allowed new
experimental investigations of two-dimension\-al mesoscopic devices
consisting in small numbers of interacting confined particles. For
instance, in type-II superconductors, SQUID microscopy~\cite{hata03}
or the multiple-small-tunnel junction method~\cite{kanda04} have
made possible the tracking of the vortices. Even more recently,
stable and metasta\-ble vortex configurations in superconducting
disks were observed using a Bitter decoration
technique~\cite{grigorieva06}. Many other systems are governed by
the same physics: an interparticle interaction, a confining
potential, and possibly a thermal activation~\cite{overview}.

In parallel with the experiments, the simplicity of the input
ingredients has allowed the development of numerical simulations to
determine both the equilibrium configurations and the dynamics of these
systems. However, when the energy levels are very close, the
uncertainties inherent in the numerical simulations make the
confrontation with the experiments necessary.

\begin{figure}
\resizebox{0.82\columnwidth}{!}{\includegraphics{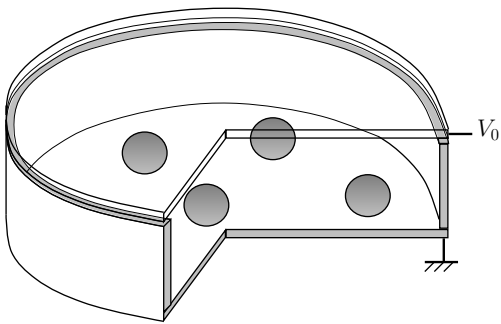}}
\caption{Section of the experimental device. Here the frame is
circular, as it will be used in section \ref{sec:equilibrium}.}
\label{fig:general}
\end{figure}

The difficulties in controlling exhaustive sets of experimental
parameters in real mesoscopic systems, such as the pinning or the
interaction strengths, have led us to devise an analogue macroscopic
experimental set-up, consisting in macroscopic Wigner islands, whose
characteristics are easily tunable. As illustrated in
Fig.~\ref{fig:general}, our macroscopic Wigner islands are
constituted by millimetric stainless steel spheres (of radius~$R =
0.4$ mm and mass~$m=2.15\, \mathrm{mg}$), sitting inside a
horizontal plane condenser of height~$h=1.5\, \mathrm{mm}$. The
bottom electrode of the condenser is a doped silicon wafer, whereas
the top one is a transparent conducting glass, thus allowing a
direct capture of the position of the spheres by means of a camera
placed above the experimental device. A metallic frame of
height~$h_c=1.5\, \mathrm{mm}$, intercalated between the two
electrodes and in electric contact with the bottom one, confines the
spheres. Note that the confining frame could be set to a different
potential than the bottom electrode, but, for simplicity, in the
following we will focus on the equality case.  When a potential
difference~$V_0$ is applied to the condenser, the spheres become
charged, repel each other and spread throughout the whole available
space, electrostatically confined by the outside frame. The
voltage~$V_0$ goes typically from a few hundred Volts to a thousand,
above which the condenser breaks down. By mechanically shaking the
whole system, we can thermalize it: the spheres acquire a brownian
motion~\cite{brownian} and obey Boltzmann statistics, in which the
shaking amplitude plays the role of an effective
temperature~\cite{coupier05}.

We then have a system in which the interaction between the spheres, as
well as the confining potential, can be easily adjusted. This device has
been previously used to study small confined
systems~\cite{stjean01,stjean02}; in such systems, as it has been shown
in numerous numerical studies~\cite{campbell79,bedanov94,lai99}, the
equilibrium positions and the dynamics are highly dependent upon the
interaction between the particles. The same experimental set-up can be
used to study pinning problems in two-dimensional elastic lattices.
Indeed, defect-free lattices of a few thousand spheres can be
obtained~\cite{stjean04}. Again, the behavior of the system, in
particular far from the equilibrium, highly depends on the interaction
potential.

The comparison of our experimental equilibrium positions with the
calculated ones, for up to thirty particles confined by a potential with
axial symmetry, had led us to think that our interparticle interaction
potential was logarithmic, at least within the experimental distance
range~\cite{stjean01}. The equilibrium configurations were also
determined with an elliptic confinement~\cite{stjean02}: the ground
state configuration for 17 particles has been compared with the one for
17 vortices resulting from the minimization of the Ginzburg-Landau free
energy in a mesoscopic type-II superconductor of the same
geometry~\cite{meyers00}. It has been suggested to the authors of the
numerical study that the configuration that they presented was not the
fundamental state but the first excited level, as they finally
confirmed~\cite{private}. Let us remind that the intervortices
interaction is described by the modified Bessel function of the second
kind $K_0$~\cite{degennes}, which has the following asymptotic
behaviors:
\begin{eqnarray}
K_0(r/\lambda)&{\sim}\atop{r\to 0}&- \ln(r/\lambda)\\
&{\sim}\atop{r\to \infty}&\sqrt{\frac{\pi \lambda}{2 r}} e^{-r/\lambda}.
\end{eqnarray}

All these indications on the nature of the interparticle interaction are
indirect evidences and depend on already existing and studied
interactions. Thus, they need to be confirmed and quantified.

In this paper, we calculate numerically this interaction by considering
the electrostatic problem of two spheres in electric contact with one of
the electrode of an infinite plane condenser.
Section~\ref{sec:interaction} is dedicated to the numerical resolution
of the problem. First, the Laplace equation is formally  solved by means
of a multipolar expansion. The numerically obtained electrostatic
potential allows to calculate the electrostatic energy of the system,
from which the interaction energy between two spheres is derived. In the
range of distances in which the numerical calculations can be made with
sufficient accuracy, the interaction energy is well fitted by a
decaying exponential as well as by a modified Bessel function~$K_0$,
both being controlled only by two parameters: their amplitude and their
screening length. In section~\ref{sec:confinement}, we determine a
simple approximation of the confining potential in the case of a
circular frame, by considering the frame as a hedge of spheres over
which the intersphere interaction energy can be integrated. A correction
of the amplitude of the confining potential must be introduced to take
into account the height difference between the interacting spheres and
the frame. This is done in section~\ref{sec:equilibrium}, where the
equilibrium configurations for up to 30 spheres and a circular frame are
calculated. Comparisons with the experimental data allow to adjust the
amplitude of the confinement and to validate the model.

\section{Interaction between two spheres}
\label{sec:interaction}
\subsection{Determination of the electrostatic potential}
\label{subsec:theory}

\begin{figure}
\resizebox{\columnwidth}{!}{\includegraphics{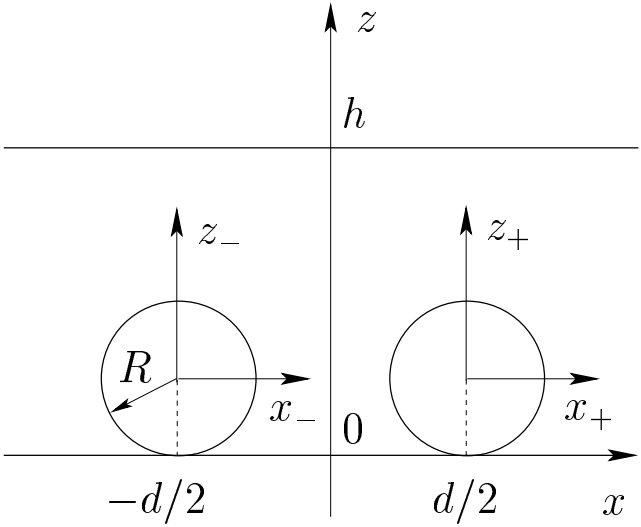}}
\caption{Geometry for the electrostatic interaction of two spheres
inside a parallel plates condenser.}
\label{fig:schema}
\end{figure}

We consider an infinite planar parallel plates condenser of
thickness~$h$; its lower plate is kept at zero potential and its upper
plate is kept at the fixed potential~$V_0$. On the lower plate, in
electric contact with it, sit two conducting spheres of radius~$R<h/2$;
the centers of the two spheres are separated by the distance~$d\ge 2R$.
We introduce a Cartesian coordinate system $\mathbf{r}=(x,y,z)$ having
the $z$-axis orthogonal to the plates of the condenser, with~$z=0$
(resp.\ $z=h$) on the lower (resp.\ upper) plate; we choose the $x$
and~$y$-axis such that the centers of the two spheres are situated
at~$x=\pm d/2$ and~$y=0$. Furthermore, we introduce two local Cartesian
coordinate systems $\mathbf{r}_-=(x_-,y_-,z_-)$
and~$\mathbf{r}_+=(x_+,y_+,z_+)$, centered, respectively, on the left
and right sphere (see Fig.~\ref{fig:schema}), and their corresponding
spherical systems of coordinates $(r_-,\theta_-,\phi_-)$
and~$(r_+,\theta_+,\phi_+)$:
\begin{eqnarray}
x_\pm &=& x\mp d/2 = r_\pm\sin\theta_\pm\cos\phi_\pm,\\
y_\pm &=& y = r_\pm\sin\theta_\pm\sin\phi_\pm,\\
z_\pm &=& z-R =  r_\pm\cos\theta_\pm.
\end{eqnarray}

To determine the interaction between the conducting spheres, we must
solve the Laplace equation
\begin{equation}
\label{eq:laplace}
\nabla^2 V = 0,
\end{equation}
for the electric potential~$V(\mathbf{r})$, with the boundary conditions
that~$V=0$ on the lower plate and on the surface of the spheres,
and~$V=V_0$ on the upper plate. To this aim, we begin to write the total
electric potential~$V(\mathbf{r})$ as the sum of the potential
$V_0\,z/h$ of the empty condenser plus a perturbation due to the
presence of the two spheres
\begin{equation}
\label{eq:V}
V(\mathbf{r}) = V_0\left[\frac{z}{h} + v(\mathbf{r})\right].
\end{equation}
The normalized perturbation~$v(\mathbf{r})$ is the solution of the Laplace
equation~(\ref{eq:laplace}) that is regular in the space inside the
condenser and outside the spheres, and that satisfies the boundary
conditions
\begin{eqnarray}
\label{eq:plates}
v(x,y,z=0) &=& v(x,y,z=h) = 0, \\
\label{eq:billeg}
v(r_-=R,\theta_-,\phi_-) &=& -\frac{z(r_-=R,\theta_-)}{h}, \\
\label{eq:billed}
v(r_+=R,\theta_+,\phi_+) &=& -\frac{z(r_+=R,\theta_+)}{h}, \\
v(|\mathbf{r}|\to\infty) &=& 0,
\end{eqnarray}
where $z(r_-=R,\theta_-)$ [resp.\ $z(r_+=R,\theta_+)$] is the
$z$-coordi\-nate of the point on the left (resp.\ right) sphere of local
spherical coordinates $(r_-=R,\theta_-,\phi_-)$ [resp.\
$(r_+=R,\theta_+,\phi_+)$]; explicitly
\begin{equation}
z(r_\pm=R,\theta_\pm) = R(1+\cos\theta_\pm).
\end{equation}

To determine the perturbation~$v(\mathbf{r})$, we start by
considering the solutions $g_{\ell m}(\mathbf{r})$ of the Laplace
equation~(\ref{eq:laplace}), regular at infinity, obtained by
separation of variables in the spherical
coordinates~$(r_-,\theta_-,\phi_-)$:
\begin{equation}
\label{eq:multi}
g_{\ell m}(\mathbf{r}) =
\frac{{P_\ell}^m (\cos\theta_-)h^{\ell+1}}{r_-^{\ell+1}}\cos(m\phi_-),
\end{equation}
where the ${P_{\ell}}^m(\cos\theta_-)$, with $\ell = 0,1,2,\ldots$ and
$0\le m\le\ell$, are the associated Legendre functions of the first
kind~\cite{abramowitz}:
\begin{equation}
{P_{\ell}}^m(t) = \frac{(-1)^{\ell+m}}{2^\ell\,\ell!}\left(1-t^2\right)^{m/2}
\frac{d^{\ell+m}}{dt^{\ell+m}}\left(1-t^2\right)^\ell.
\end{equation}
The elementary solutions~(\ref{eq:multi}) correspond to the usual terms
of the expansion of the electrostatic potential in multipolar
moments~\cite{landau}; they are singular only at the center of the left
sphere, $r_-=0$. The factors $h^{\ell+1}$ in their definition have been
introduced to make them dimensionless. Note also that we retained only
the harmonics~$\cos(m\phi_-)$, that satisfy the symmetry with respect to
the $(x,z)$-plane of the present geometry. To construct, starting from
the multipoles~(\ref{eq:multi}), a suitable complete basis of harmonic
functions for our situation, we make a set of infinite mirror images of
the potentials~(\ref{eq:multi}) through the two plates of the condenser
$z=0$ and~$z=h$, such that the boundary conditions~(\ref{eq:plates}) are
automatically satisfied, and we symmetrize the resulting potential
through the $(y,z)$-plane, as required by our geometry. The resulting
base functions are therefore
\begin{eqnarray}
\label{eq:basis}
&&G_{\ell m}(\mathbf{r}) = \sum_{k=-\infty}^\infty\Big[
g_{\ell m}(x,y,z+2kh)\nonumber\\
&&-g_{\ell m}(x,y,-z+2kh)+g_{\ell m}(-x,y,z+2kh)\nonumber\\
&&-g_{\ell m}(-x,y,-z+2kh)\Big];
\label{eq:base}
\end{eqnarray}
they are regular in the volume inside the condenser and outside the
spheres, since their only singularities are at the centers of the
spheres and at their successive mirror images through the two plates of
the condenser, \textit{i.e.}, at $(x=\pm d/2,y=0,z=\pm R + 2kh)$, with
$k=0,\pm 1,\pm 2,\ldots$. Using this set of base functions, we decompose
the perturbation~$v(\mathbf{r})$ as
\begin{equation}
\label{eq:dec}
v(\mathbf{r}) = \sum_{\ell=0}^\infty \sum_{m=0}^\ell v_{\ell m} G_{\ell
m} (\mathbf{r}).
\end{equation}
The unknown coefficients~$v_{\ell m}$ depend on the distance $d$
between the two spheres and can be determined by imposing the
remaining boundary conditions (\ref{eq:billeg})
and~(\ref{eq:billed}) on the two spheres. Since the basis
functions~(\ref{eq:basis}) are symmetric with respect to
the~$(y,z)$-plane, only one of the two conditions has to be imposed.
This is most efficiently done by projecting the boundary
condition~(\ref{eq:billeg}) on the set of functions
${P_{\ell'}}^{m'} (\cos\theta_-) \cos(m'\phi_-)$, that represents a
complete set of orthogonal functions---with the required symmetry
with respect to the $(x,z)$ plane---on the left sphere. The
resulting boundary equations are therefore
\begin{eqnarray}
\label{eq:vlm}
\sum_{\ell=0}^\infty \sum_{m=0}^\ell v_{\ell m}\!\!
&&\int_0^\pi\!\!\! \sin\theta_- d\theta_- \!\int_0^{2\pi}\!\!\! d\phi_-
\Big[G_{\ell m} (r_-\!=\!R,\theta_-,\phi_-)\times\nonumber\\
&&{P_{\ell'}}^{m'} (\cos\theta_-) \cos(m'\phi_-)\Big]=\nonumber\\
&&\left\{
\matrix{-\displaystyle\frac{4\pi R}{h}& \mbox{if $\ell'=m'=0$}\hfill\cr
&\cr
-\displaystyle\frac{4\pi R}{3h}& \mbox{if $\ell'=1$ and $m'=0$}\hfill\cr
&\cr
0& \mbox{otherwise}\hfill\cr} \right. .
\end{eqnarray}

To numerically determine the coefficients~$v_{\ell m}$, we
truncate the sum in the left-hand side of Eq.~(\ref{eq:vlm}) to a
finite $\ell = \ell_m$, and we evaluate Eq.~(\ref{eq:vlm}) for $m'=0, 1,
\ldots, \ell'$ and~$\ell'=0,1, \ldots,\ell_m$. This gives a set of
$(\ell_m+1)(\ell_m+2)/2$ linear equations in the
$(\ell_m+1)(\ell_m+2)/2$ unknowns~$v_{\ell m}$. The integrals in
Eq.~(\ref{eq:vlm}), that determine the coefficients of this linear
system, are computed numerically by truncating the mirror images sum in
Eq.~(\ref{eq:basis}) to a finite $|k|=k_m$. By varying $\ell_m$
and~$k_m$, we check for the convergence of the expansion.

\subsection{Interaction energy} \label{subsec:interaction}

\begin{figure}
\resizebox{\columnwidth}{!}{\includegraphics{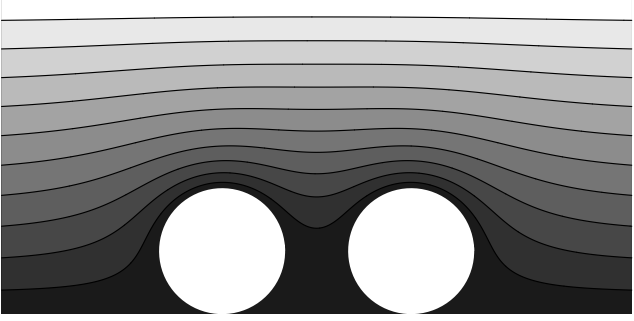}}
\caption{Contour plot of the electrostatic potential~$V(\mathbf{r})$ in
the plane~$y=0$ for two spheres of radius~$R=0.2\,h$ at the
center to center distance $d=0.6\, h$. The spheres are indicated in
white. Darker shadings correspond to lower values of~$V(\mathbf{r})$.}
\label{fig:xz}
\end{figure}

\begin{figure}
\resizebox{\columnwidth}{!}{\includegraphics{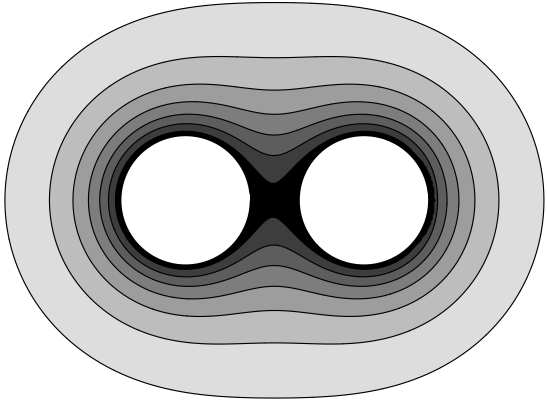}}
\caption{Same as figure~\protect\ref{fig:xz} but in the plane $z=R$,
\textit{i.e.}, the plane parallel to the plates of the condenser and
passing through the centers of the spheres.}
\label{fig:xy}
\end{figure}

At fixed potential difference~$V_0$, the mechanical interaction
energy~$F$ between the spheres is equal to the opposite of the
electrostatic energy stored in the condenser
\begin{equation}
\label{eq:mec}
F = -\frac{1}{2} Q V_0,
\end{equation}
where~$Q$ is the charge stored on the upper plate~$z=h$. The latter, for
our infinite geometry, is actually infinite. However, its
variation~$\Delta Q$ due to the introduction of the spheres is finite:
this is the charge associated to the potential perturbation~$V_0\,
v(\mathbf{r})$ in Eq.~(\ref{eq:V}). To compute it, we make use of the
reciprocity theorem, according to which, given two charge distributions
$\rho(\mathbf{r})$ and~$\rho'(\mathbf{r})$ that create the potentials
$V(\mathbf{r})$ and~$V'(\mathbf{r})$, respectively,
\begin{equation}
\label{eq:reciproc}
\int \rho(\mathbf{r})\, V'(\mathbf{r})\, d\mathbf{r} =
\int \rho'(\mathbf{r})\, V(\mathbf{r})\, d\mathbf{r},
\end{equation}
the volume integrals being performed over all the space~\cite{durand}.
As $(\rho',V')$ system, we consider a plane parallel condenser having
its lower plate~$z=0$ at zero potential and upper plate~$z=h$ at the
potential~$V_0$; its potential distribution is thus $V'=V_0 z/h$ for
$0\le z \le h$, $V'=0$ for $z\le 0$ and~$V'=V_0$ for~$z\ge h$. As
$(\rho,V)$ system we consider the potential distribution having $V=V_0\,
v(\mathbf{r})$ for $0\le z\le h$ and~$V=0$ otherwise; its associated
charge distribution has three contributions: a surface charge
$\Delta\sigma_-(x,y)$ on the lower plate $z=0$, a surface charge
$\Delta\sigma_+(x,y)$ on the upper plate~$z=h$, and two point-like
charge distributions centered on the two spheres, at~$(x=\pm d/2,
y=0,z=R)$, associated to the corresponding multipolar singularities of
the perturbation~$v(\mathbf{r})$. The surface integral of the charge
distribution~$\Delta\sigma_+(x,y)$ is equal to the charge
variation~$\Delta Q$ due to the introduction of the spheres inside the
condenser:
\begin{equation}
\label{eq:dsigma}
\int_{z=h} \Delta\sigma_+(x,y) \, dx\, dy = \Delta Q.
\end{equation}
Now:
\begin{equation}
\int \rho'(\mathbf{r})\, V(\mathbf{r})\, d\mathbf{r} = 0,
\end{equation}
since $\rho'$ is non-zero only for $z=0$ and~$z=h$, where~$V$ is zero.
On the other hand,
\begin{equation}
\label{eq:rhovp}
\int\!\! \rho(\mathbf{r})\, V'(\mathbf{r})\, d\mathbf{r} =
V_0 \int_{z=h}\!\!\!\! \Delta\sigma_+(x,y) \, dx\, dy
+ \frac{V_0}{h} \int_{\Omega_R}\!\! \rho(\mathbf{r})\, z \,d\mathbf{r}.
\end{equation}
The last volume integral in this expression is performed in the region
$\Omega_R$ made of two arbitrary volumes encircling the two centers of the
spheres, where $\rho(\mathbf{r})$ is concentrated: it equals the
$z$-component of the dipolar moment of the charge
distribution~$\rho(\mathbf{r})$ around the
origin~$\mathbf{r}=\mathbf{0}$. Using Poisson's equation
$\rho(\mathbf{r}) = -\epsilon_0 \nabla^2 [V_0 v(\mathbf{r})]$, where
$\epsilon_0$ is the vacuum dielectric constant, by successive
integrations by parts and application of the Green theorem, we obtain
\begin{equation}
\int_{\Omega_R}\!\! \rho(\mathbf{r})\, z \,d\mathbf{r} =
\epsilon_0 V_0\oint_{\partial\Omega_R}\left[ v(\mathbf{r})
\mathbf{z}\cdot\bm{\nu} - z\frac{\partial v}{\partial\nu}
\right]\, dS,
\end{equation}
where the surface integral is performed over two arbitrary surfaces
$\partial\Omega_R$ of outward normal~$\bm{\nu}$ around the two centers
of the spheres and $\mathbf{z}$ is the unit normal in the $z$-direction.
Taking for $\partial\Omega_R$ two spherical surfaces centered at $(x=\pm
d/2, y=0, z=R)$ and using the decomposition~(\ref{eq:dec}), one obtains
\begin{equation}
\label{eq:dip}
\int_{\Omega_R}\!\! \rho(\mathbf{r})\, z \,d\mathbf{r} = 8\pi\epsilon_0
V_0 h^2\left[v_{00} \frac{R}{h} +v_{10}\right].
\end{equation}
In fact, the only terms in the multipoles~(\ref{eq:multi}) having a
non-zero $z$-component of the dipole moment with respect to the
point~$\mathbf{r}=\mathbf{0}$ are the monopole (charge) term $\ell=m=0$
and the dipole term~$\ell=1$, $m=0$. Finally, putting together
Eqs.\ (\ref{eq:reciproc})--(\ref{eq:rhovp}) and~(\ref{eq:dip}), the
charge variation due to the introduction of the spheres at constant
potential~$V_0$ can be expressed as
\begin{equation}
\Delta Q = -8\pi\epsilon_0 V_0 h \left[v_{00} \frac{R}{h}
+v_{10}\right].
\end{equation}
From Eq.~(\ref{eq:mec}), the variation of the mechanical energy of the
spheres is then
\begin{equation}
\label{eq:Deltaf}
\Delta F(d) = \epsilon_0 V_0^2 h\, f(d),
\end{equation}
with the normalized mechanical energy
\begin{equation}
\label{eq:fint}
f(d) = 4\pi\left[v_{00} \frac{R}{h}+v_{10}\right].
\end{equation}

\subsection{Numerical results} \label{subsec:numresult}

In Figs.\ \ref{fig:xz} and~\ref{fig:xy} we show typical
cartographies of the electrostatic potential. As it is apparent from
Fig.~\ref{fig:xz}, above the top of each sphere the equipotentials
are uniformly squeezed over the whole remaining thickness. Thus the
wavelength of the corresponding perturbation along~$z$ is comparable
with the thickness~$h$ of the condenser. Since $\nabla^2 V=0$, this
implies a lateral relaxation on the same length scale: we thus
expect that the interaction between the spheres is short ranged with
a decay length~$\approx h$ (rather than~$R$).

\begin{figure}
\resizebox{\columnwidth}{!}{\includegraphics{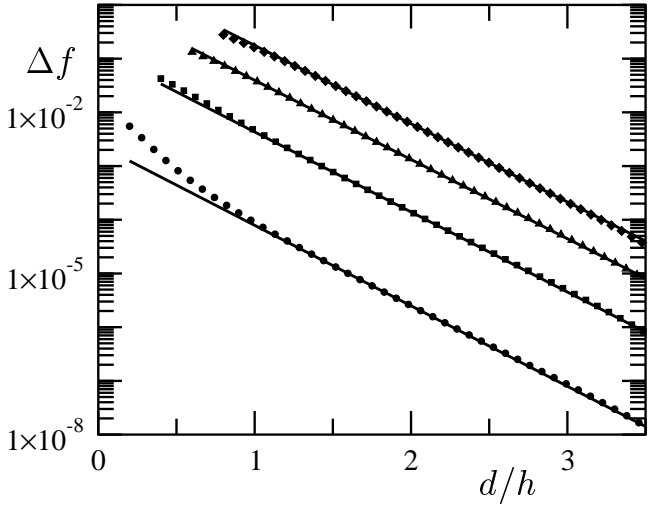}}
\caption{Normalized mechanical interaction energy~$\Delta f$ as a
function of the normalized distance $d/h$ between the centers of the
two spheres for 4 different radii of the spheres: $R=0.1\, h$
(dots), $R=0.2\, h$ (squares), $R=0.3\, h$ (triangles), $R=0.4\, h$
(diamonds). The lines indicate the exponential fit of the numerical
data.} \label{fig:inter}
\end{figure}

\begin{figure}
\resizebox{\columnwidth}{!}{\includegraphics{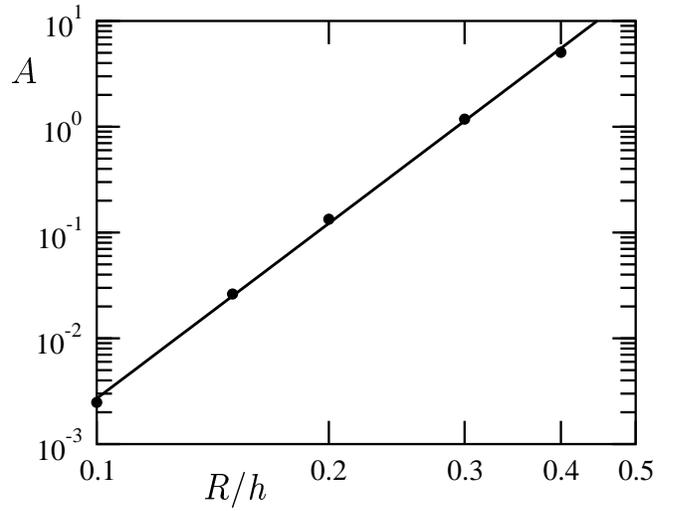}}
\caption{Amplitude~$A$ of the exponential fit of the mechanical
interaction energy (see Eq.~\protect\ref{eq:fit}) as a function of
the normalized radius~$R/h$ of the two spheres. Dots: numerical
data. Continuous line: power-like fit~$A = 852\,(R/h)^{5.5}$.}
\label{fig:amplitude}
\end{figure}

In Fig.~\ref{fig:inter} we present the numerically computed mechanical
interaction energies $\Delta f(d) = f(d)-f(\infty)$ as a function of the
distance~$d$ between the two spheres, for different radii~$R$ of the
spheres, from contact up to a distance equal to $3.5$ times the
thickness of the condenser. Over this range of distances and for spheres
of intermediate size, roughly filling half of the height of the
condenser (triangles in Fig.~\ref{fig:inter}, corresponding
to~$R=0.3\,h$), the numerical data are quite well fitted by an exponential
law
\begin{equation}
\label{eq:fit}
\Delta f(d) = A \exp(-d/\lambda),
\end{equation}
with the decay length~$\lambda\simeq 0.29 h$. For smaller or larger
spheres, this exponential behavior is violated at small distances,
while it is recovered at distances larger than~$h$, with a decay
length essentially independent of the size of the spheres. The
amplitude~$A$ of the interaction grows with the radius of the
spheres; as shown in Fig.~\ref{fig:amplitude}, it can reasonably
well approximated by the power-like behavior~$A \simeq
852\,(R/h)^{5.5}$.

\begin{figure}
\resizebox{\columnwidth}{!}{\includegraphics{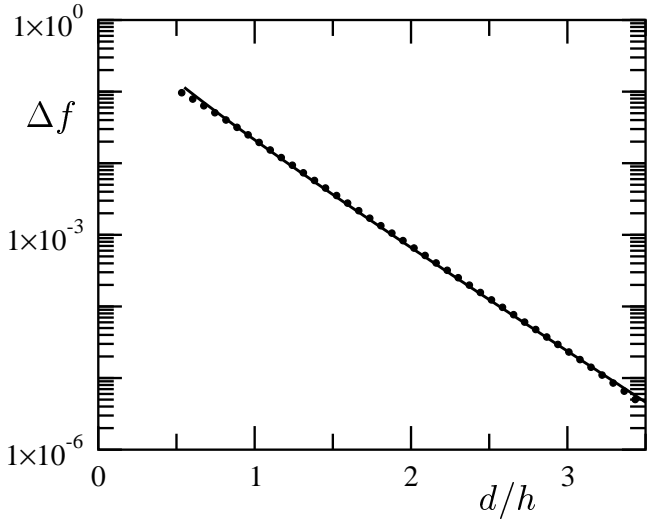}}
\caption{Normalized mechanical interaction energy~$\Delta f$ as a
function of the normalized distance $d/h$ between the centers of the
two spheres for the experimental radius $R=0.267\, h$ of the spheres.
Dots: numerical results. Continuous line: fit of the numerical data with
the modified Bessel function~$K_0$.}
\label{fig:fitk0}
\end{figure}

\begin{figure}
\resizebox{\columnwidth}{!}{\includegraphics{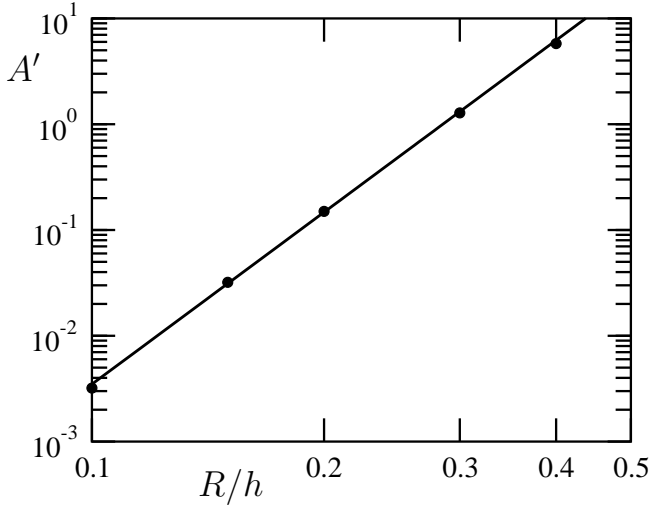}}
\caption{Amplitude~$A'$ of the~$K_0$ fit of the mechanical
interaction energy (see Eq.~\protect\ref{eq:fitk0}) as a function of
the normalized radius~$R/h$ of the two spheres. Dots: numerical
data. Continuous line: power-like fit~$A' = 876\,(R/h)^{5.4}$.}
\label{fig:ampk0}
\end{figure}

Nevertheless, we find that within the explored range for
the distance $d$, which corresponds to the experimental situations,
the numerical data are also well fitted by a~$K_0$ law
\begin{equation}
\label{eq:fitk0} \Delta f(d) = A' K_0(d/\lambda'),
\end{equation}
as it is shown in Fig.~\ref{fig:fitk0} for the radius $R=0.267\, h$
corresponding to the real radius of the spheres in the experimental
device. The decay length is then slightly higher, $\lambda'\simeq
0.32\,h$; the behavior of the corresponding amplitude~$A'$ as a function
of the size of the spheres is shown in Fig.~\ref{fig:ampk0}: again, it
is well fitted by a power-like behavior~$A'\simeq 876\,(R/h)^{5.4}$ with
essentially the same exponent. Indeed, the presence of a sphere inside
the condenser introduces a perturbation of the electrostatic potential
that can be expanded in Fourier harmonics along~$z$. The first harmonic
varies along~$z$ as $\sin(\pi z/h)$: far from the axis of the sphere,
the corresponding solution of the Laplace equation that has cylindrical
symmetry along its axis and that decays to zero at infinity is
proportional to $\sin(\pi z/h) K_0(\pi d/h)$, where~$d$ is the distance
from the axis of the particle.  We thus expect~$\lambda=h/\pi \simeq
0.318\,h$.

\section{Interaction with a confining ring}
\label{sec:confinement}

In our experiment, the spheres are laterally confined by a frame that is
in electric contact with the lower plate and almost touches the upper
one. Here, we focus on the case of a circular ring of radius~$R_c$ and
height~$h_c$. To treat this confinement using the results of the
preceding section, we model it as a necklace of touching spheres of
diameter~$h_c$, whose centers lie on a circle of radius $R_c$. We make
the additional simplifying assumption that the electrostatic
interactions are pairwise additive. These two hypotheses are justified
for distances to the confining ring large with respect to its height.

For simplicity, as normalized interaction energy~$v_{c0}$ between a
sphere and one of the fictitious spheres of the confining ring we take
the exponential approximation
\begin{equation}
v_{c0} = v_0 \exp(-d/\lambda),
\end{equation}
where the amplitude~$v_0$ depends on the radius of the sphere and on the
height of the ring. The center to center distance~$d$ can be expressed
as $d = (r^2+R_c^2-2 r R_c \cos\phi)^{1/2}$, as a function of the
distance~$r$ between the center of the sphere and the center of the
confining ring, and the angle~$\phi$ between the two spheres, seen from
the center of the ring. The screening length $\lambda$, which does not
depend on the radius of the spheres, is the same as the one determined
in subsection~\ref{subsec:numresult}. The total normalized confining
potential can then be written as
\begin{equation}
\label{eq:confd}
v_c = v_0 \sum_{n=1}^M \exp\left[-\lambda^{-1}\sqrt{r^2+R_c^2-2 r R_c
\cos\left(\frac{2\pi n}{M}\right)}\right],
\end{equation}
where $M=2\pi R_c/h_c$ is the number of the necklace spheres making up
the confining ring. To obtain a simpler analytical expression, we
approximate the sum by an integral. This is well justified in the
limit~$R_c\gg h_c$, which is already implicit in the hypothesis that the
distance~$R_c-r$ at the confining ring is large with respect to its
height~$h_c$. In this limit, the confining potential becomes
\begin{equation}
v_c \simeq \frac{M v_0}{2\pi} \int_0^{2\pi}\!\!\!\! \exp\left[-\lambda^{-1}
\sqrt{r^2+R_c^2-2 r R_c \cos\phi}\right]d\phi.
\end{equation}
This integral cannot be expressed analytically in terms of elementary
functions.  However, a reasonably good approximation for $r\simeq R_c/2$
can be obtained by the saddle-point method~\cite{mathews}
\begin{equation}
\label{eq:confa}
v_c(r \simeq R_c/2) \simeq M v_0
\sqrt{\frac{\lambda}{2\pi}\left(\frac{1}{r}-\frac{1}{R_c}\right)}
\exp\left[\frac{r-R_c}{\lambda}\right].
\end{equation}
Finally, as $R_c\gg \lambda$ and since it is found numerically in
that case that the logarithm of the confining
potential~(\ref{eq:confd}) has a dependence on~$r$ that is not far
from linear, we approximate it with the first-order Taylor expansion
of the logarithm of the approximate potential~(\ref{eq:confa})
around $r=R_c/2$. The resulting approximate confining potential that
we shall use in the following is then
\begin{equation}
\label{eq:confining}
v_c(r) \simeq v_0\frac{\sqrt{2 \pi\lambda R_c}}{h_c}
\exp{\left[1 - \frac{2 r}{R_c} + \frac{r - R_c}{\lambda }\right]}.
\end{equation}
This expression fits very well the discrete sum~(\ref{eq:confd}) in the
whole range of validity of the latter. Note that $v_0$ still needs to be
determined, which will be done in the following section, where the
approximations made above will be validated.

\section{Equilibrium configurations}
\label{sec:equilibrium}

\begin{table}
\begin{tabular}{ccc|ccc}
$N$&Experiments&Theory&$N$&Experiments&Theory\\
\hline\hline
5&5&5&18&1-6-11&1-6-11\\
6&1-5&1-5&19&1-6-12&1-6-12\\
7&1-6&1-6&20&1-6-13&1-6-13\\
8&1-7&1-7&21&1-7-13&1-7-13\\
9&1-8&1-8&22&1-7-14&1-7-14\\
10&2-8&2-8&\textbf{23}&\textbf{2-8-13}&\textbf{1-8-14}\\
11&3-8&3-8&24&2-8-14&2-8-14\\
12&3-9&3-9&25&3-8-14&3-8-14\\
13&4-9&4-9&26&3-9-14&3-9-14\\
14&4-10&4-10&27&3-9-15&3-9-15\\
15&4-11&4-11&28&3-9-16&3-9-16\\
16&5-11&5-11&\textbf{29}&\textbf{4-9-16}&\textbf{4-10-15}\\
17&1-5-11&1-5-11&\textbf{30}&\textbf{4-9-17}&\textbf{4-10-16}\\
\hline\hline
\end{tabular}
\caption{\label{tab:configurations} Experimental and theoretical ground
state configurations for $5\le N\le 30$ spheres of radius $R=0.4\,
\mathrm{mm}$ in a condenser of height~$h=1.5\, \mathrm{mm}$ confined
inside a ring of radius $R_c=5\, \mathrm{mm}$. The cases were there is
a discrepancy between the experimental observation and the theoretical
prediction are set in boldface.}
\end{table}

In order to validate our model, we consider the equilibrium states
of small Wigner islands. More precisely, we will successively focus
on the general configurations, then on the precise positions of the
spheres and finally on the energetic differences between the stable
and metastable states.

We use the approximate interparticle and confining potentials
(\ref{eq:fit}) and~(\ref{eq:confining}) to determine numerically the
equilibrium configurations of our macroscopic Wigner islands. The
theoretical predictions are compared with the experimental
observations, as it was done in Ref.~\cite{stjean01} for some model
interaction potentials. We recall that our Wigner islands consist
of~$5\le N\le 30$ spheres of radius~$R=0.4\, \mathrm{mm}$ contained
in a condenser of height~$h=1.5\, \mathrm{mm}$ held at the potential
difference~$V_0=700\,\mathrm{V}$; the spheres are confined within a
disc of radius~$R_c=5\, \mathrm{mm}$ and height~$h_c$ practically
coinciding with~$h$.

The theoretical stable and metastable equilibrium configurations are
obtained by searching numerically for the local minima of the total,
pairwise additive, normalized interaction energy
\begin{equation}
\label{eq:H}
H=\sum_{1\le i<j\le N}
\Delta f (|\mathbf{r}_i-\mathbf{r}_j|)+\sum_{1\le i\le N}
v_c(|\mathbf{r}_i|),
\end{equation}
where $\mathbf{r}_i$ is the 2D position of the $i$-th~sphere with
respect to the center of the ring. According to
Eq.~(\ref{eq:Deltaf}), the interaction energy is normalized with
respect to~$\epsilon_0 V_0^2 h$. To minimize~$H$, we employ a
conjugate gradient method~\cite{conjugate}, starting from many
suitably chosen different initial conditions, in order to explore a
significant portion of the complex energy landscape and find the
various relative minima.

The stable and metastable configurations form patterns roughly
constituted by concentric shells on which the spheres are located. We
shall refer to such patterns by means of the
notation~$(N_0$-$N_1$-$N_2$-$\ldots)$, where $N_i$ is the number of
spheres in the $i$-th shell from the center.

\begin{figure}[t]
\resizebox{\columnwidth}{!}{\includegraphics{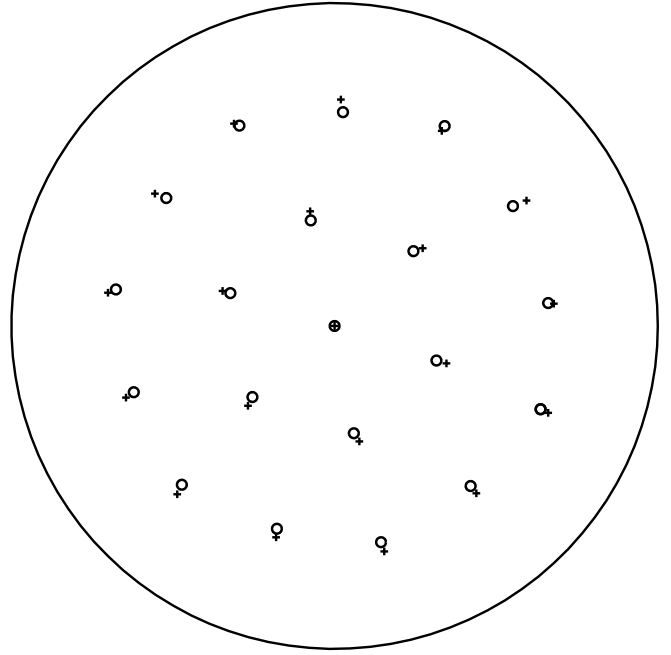}}
\caption{Ground state configuration for $N=20$ spheres. The external
circle is the confining ring, the small circles represent the
experimental positions of the spheres, drawn to scale. The plus are the
center of the theoretical positions of the spheres.}
\label{fig:conf}
\end{figure}

The total interaction energy~(\ref{eq:H}) depends on three parameters:
the decay length~$\lambda$, the amplitude of the interparticle
interaction~$A$ [see Eq.~(\ref{eq:fit})], and the amplitude of the
interaction with the ring~$v_0$ [see Eq.~(\ref{eq:confining})]. We take
the exponential approximation with $\lambda=0.29h$ and~$A=0.68$ as
determined by a best fit of the numerical interaction energy between two
equal spheres of radius~$R=0.267h$. The remaining parameter~$v_0$, that
depends on the height of the confining ring, is determined by adjusting
the radius of the ground state configuration for~$5$ spheres, that
consists in a single shell of 5 particles, to the experimental
radius~$r=2.25\, \mathrm{mm}$. We thus obtain~$v_0=0.47$. We note that,
contrary to what one would expect on the basis that~$h_c> 2R$
(\textit{i.e.}, the necklace confining spheres are larger than the
interacting spheres), we have~$v_0<A$. However, it must be noted that,
as we pointed out, the approximate confining interaction energies
(\ref{eq:confd}) and~(\ref{eq:confining}) are justified only in the
limit~$R_c\gg h_c$, a condition that is not satisfied in our
experimental situation, where~$R_c/h_c \simeq 3$. In that conditions,
trying to determine more precisely~$v_0$ by studying the interaction of
two spheres of unequal heights would be pointless, since the
approximation of the confining frame as a wedge of spheres would still
remain. Nonetheless, as we shall see, our approximate interaction
energies reasonably well account for almost all our experimental
observations.

\subsection{Ground state configurations}

In Table~\ref{tab:configurations} we report a comparison between the
experimental and the theoretical ground state configurations for a
number~$N$ of spheres ranging from $5$ to~$30$. All but three of the
ground state configurations are correctly predicted. In
Ref.~\cite{stjean01}, these same configurations were compared with
predictions derived from other model interaction potentials: the
best agreement was found for a logarithmic potential, that fails to
predict the correct ground state only in 4 cases, the same 3 cases
as our present model ($N=23,29,30$), plus the $N=28$ situation.

As for the previously studied interactions, the observed
discrepancies, that occur for dense packing of the spheres, are
irrelevant, mainly because of the extreme smallness of the relative
energy difference between the ground and the first excited state, as
for the $N=23$ and $N=29$ cases. Moreover, for dense packing the
hypothesis of pairwise interaction could be too strong and steric
effects might make the experimental determination of the ground
state tougher. Let's mention finally that another possible source of
error might be the incorrect treatment of the confining potential
for spheres too close to the ring.

\subsection{Relative positions in the configurations}

In Fig.~\ref{fig:conf} we compare the experimental positions of the
spheres in the ground state configuration for $N=20$ spheres with
the predicted one. We stress that these results are not a fit of the
experimental data, the only adjustable parameter of our model, the
relative amplitude of the confining potential~$v_0$ appearing in
Eq.~(\ref{eq:confining}) having been set once for all by comparison
with the situation for $N=5$~spheres. Again, given that the
condition of validity $R_c/h_c\gg 1$ of our approximate confining
potential is not satisfied, the agreement between the experimental
data and our simple model is rather encouraging. The same good
agreement is found for all the ground and excited state
configurations. In particular, in Table~\ref{tab:radii} we compare
the experimental and theoretical radii of the first two states for
$N=5$ and~$N=6$ spheres, along with the analytical prediction for a
logarithmic potential with a parabolic confinement. This confirms
that our theoretical interaction is closer to the real one than the
previously predicted ones.

\begin{table}
\begin{center}
\begin{tabular}{c|c|c|c}
Configuration& $R_\mathrm{exp}\, [\mathrm{mm}]$ &
$R_\mathrm{th}\, [\mathrm{mm}]$ & $R_\mathrm{log}\, [\mathrm{mm}]$\\
\hline\hline
5&$2.25$ & $2.25$ & 2.25$$\\
1-4&$2.44$ & $2.45$ & 2.52$$\\
1-5&$2.48$ & $2.50$ & 2.76$$\\
6&$2.35$ & $2.43$ & 2.52$$\\
\hline\hline
\end{tabular}
\end{center}
\caption{\label{tab:radii} Radii of the outer shell of 4 different
stable (5 and 1-5) and metastable (1-4 and 6) configurations.
$R_\mathrm{exp}$: experimental values; $R_\mathrm{th}$: theoretical
values according to our model; $R_\mathrm{log}$: theoretical value
for a logarithmic potential with a parabolic confining. The two
theoretical values were obtained by tuning the confining potential
such as to reproduce the experimental observation corresponding to
the first configuration.}
\end{table}

\subsection {Energy levels}

Finally, the energy differences between the ground and the first
excited states have been explored. Experimentally, the possibility
to submit the system, through a mechanical shaking, to an effective
temperature $T$ that has already been calibrated allows us to
explore these different states~\cite{coupier05}. Neglecting the
higher excited levels, that have a much larger energy, the
difference $\Delta E$ between the energy~$E_2$ of the first excited
level and the energy~$E_1$ of the ground state is then obtained by
measuring the ratio of the respective mean residence times
$<\tau_2>$ and~$<\tau_1>$~\cite{kramers40}:
\begin{equation}
\label{kramers} \frac{<\tau_1>}{<\tau_2>}\propto\mathrm{e}^{\Delta
E/k_B T}.
\end{equation}

Table~\ref{tab:energy} reports a comparison between the experimental
and theoretical excitation energies for $N=18$, $19$, and~$20$
spheres. We also report the theoretical absolute values of the
interaction energies of the first excited states. The agreement
between the theoretical predictions and the experimental data is
qualitatively good, considering that the relative differences
$\Delta E/E_2$ are extremely small. This last comparison finally
validate our model.

\section{Conclusion}
\label{sec:conclusion}

Using a semi-analytical method, we have performed a direct calculation
of the interaction between two conducting spheres lying on the bottom
electrode of a plane condenser. We find that, within a significant
range, the interaction energy can be described by a simple decaying
exponential, as well as by a~$K_0$ function, both being governed by a
screening length that is typically a third of the condenser's height. On
the basis of this interaction, our theoretical predictions for small
Wigner islands constituted by up to a few tens of spheres are in good
accordance with the experimental observations, thus validating our
simple model.

This study definitively completes the description of our system of
interacting spheres, where the density, the interaction amplitude and
the temperature are now well-known and can easily be tuned. Thus, this
system is a macroscopic experimental model that easily allows to explore
the properties of two-dimensional confined systems, such as vortices in
mesoscopic type-II superconductors. In particular, the understanding of
the dynamics of the vortices could be enriched in a complementary way by
corresponding studies on the macroscopic Wigner islands.

\begin{table}
\begin{center}
\begin{tabular}{c|c|c|c|c}
$N$&Excited state& $E_\mathrm{th}\, [\mathrm{J}]$ &
$\Delta E_\mathrm{th}\, [\mathrm{J}]$& $\Delta E_\mathrm{exp}\,[\mathrm{J}]$\\
\hline\hline
18 & 1-5-12 & $4.1\times 10^{-9}$ & $1.8\times 10^{-11}$ &
$4.8\times 10^{-11}$\\
19 & 1-7-11 &$4.7\times 10^{-9}$ & $0.9\times 10^{-10}$ &
$1.3\times 10^{-10}$\\
20 & 1-7-12 &$5.3\times 10^{-9}$ & $0.3 \times 10^{-11}$ &
$1.5 \times 10^{-11}$\\
\hline\hline
\end{tabular}
\end{center}
\caption{\label{tab:energy} Theoretical energies~$E_\mathrm{th}$ of the
first excited state for three different numbers~$N$ of spheres and
theoretical~$\Delta E_\mathrm{th}$ and experimental~$\Delta
E_\mathrm{exp}$ differences of the energy of the excited state with
respect to the ground state.}
\end{table}


\begin{thebibliography}{10}
\bibitem{overview}
Other related systems are, for instance, colloids, electrons in
quantum dots,
  vortices in superfluid $^4\mbox{He}$, electron dimples on a liquid helium
  surface, trapped cooled ions or vortices in a Bose-Einstein condensate.
\end{thebibliography}
\end{document}